\documentclass{mem}
\usepackage{natbib}\usepackage{txfonts}\usepackage{balance}
\usepackage{graphicx}
\usepackage[a4paper]{hyperref}
\idline{75}{282}
\begin{document}
\def\teff{$T\rm_{eff }$}
\def\kms{$\mathrm {km s}^{-1}$}
\newcommand{\lsp}{LS~I~+61$^{\circ}$303}
\newcommand{\grs}{GRS 1915+105~}
\newcommand{\lsi}{LS~I~+61$^{\circ}$303~}
\newcommand{\ls}{LS~5039}

\title{
The Two-Peak Model  of LS I +61303: 
Radio Spectral Index Analysis 
}

   \subtitle{}

\author{
M. \,Massi\inst{1} }

% \offprints{M. Massi}

\institute{
Max Planck Institut f\"r Radioastronomie,
Auf dem H\"ugel 69,
D-53121, Germany\\
\email{mmassi@mpifr-bonn.mpg.de}
}

\authorrunning{Massi }

\titlerunning{Steady Jets and Transient Jets}

\abstract{
The most puzzling aspect of the radio emission from \lsi
is  that the  large periodic radio outburst, with period equal to the orbital one, 
occurs very displaced from periastron passage, nearly at apoastron.
In 1992,  Taylor, one of the discoverers of this source, 
together with his collaborators proposed 
a model of a compact object in an eccentric orbit
accreting from the equatorial wind of the Be star primary. 
The application of this model by
\citet{martiparedes95} predicts
one ejection at periastron  and a second more displaced ejection along the orbit.
The first ejection should correspond to weak radio emission, because of strong inverse
Compton losses of the emitting electrons due to the proximity to
the hot Be star, 
whereas  the second ejection, quite  displaced from the star, would correspond to a strong radio outburst, that one indeed  observed.
Corroborated along the years by numerical computations, 
simulations and gamma-ray observations, until now  this two-peak model could not be proved  
in the radio band, because of the negligible emission around periastron.
We show here, that  the radio spectral index based on the  ratio of flux densities 
is the unique tool to monitor activity of \lsi 
in the radio band around periastron.
The  analysis of the radio spectral index over almost 7 years of Green Bank Interferometer
 data results in  a  clear double-peaked 
spectral index curve 
along the orbit. 
This result  gives finally  observational support 
at radiowavelengths to  the two-peak accretion/ejection model for \lsp.  
Moreover, the here shown comparison of the two-peak curves - the radio spectral index curve and the Fermi-LAT gamma-ray curve - indicates a new interesting 
hypothesis on  the electron population responsible for the gamma-ray emission.
\keywords {Radio continuum: stars -- X-rays: binaries -- gamma-rays: observations -- X-rays: individual: LSI+61303}
}
\maketitle{}

\section{Introduction}
The TeV-emitting source \lsi is a X-ray binary system where a compact object
travels through the dense equatorial wind of a Be star.
The most typical peculiarity of \lsi is a large periodic radio outburst 
toward apoastron.
In Fig. 1 we see 6.7yr of Green Bank Interferometer data
folded with the orbital period  $P_1$=26.496 d 
 ($\Phi$:  ${(t-t_0)}\over P_1$,
with t$_0$=JD\,2443366.775)
\citep{gregory2002}. 
Periastron passage corresponds to orbital phase $\Phi$=0.23 for 
\citet{casares05}  and to $\Phi=0.275$ for \citet{aragona09}. 
As one can see in  Fig. 1,  the outburst is clustered around $\Phi\simeq$0.6, i.e. 
almost apoastron.
The broad shape of the light curve of Fig.1  is due to variations of the orbital phase and amplitude of the outburst, both
changing with a long-term  period 
$P_2$=1667~d
 \citep{gregory2002}
 ($\Theta$:  ${(t-t_0)}\over P_2$).
% This long term  variation is 
%attributed  to changes in the 
%parameters (density, velocity) of the wind  of the Be donor star \citep{zamanovmarti00}. 
\citet{zamanovmarti00} demonstrated a modulation on the same timescale in the H$_\alpha$ emission line.
This latter result strongly suggests
that the long-term modulation in the radio emission is related to
changes in the Be star equatorial disk properties.
In particular, \citet{gregoryneish02} suggest  that the
long-term  modulation in radio properties may stem from 
periodic
ejections of a shell (density enhancement) of gas in the
equatorial disk of the Be star. 

\section{Two Peak Accretion Model}
\begin{figure}
\resizebox{\hsize}{!}{\includegraphics[clip=true, angle=-90.]{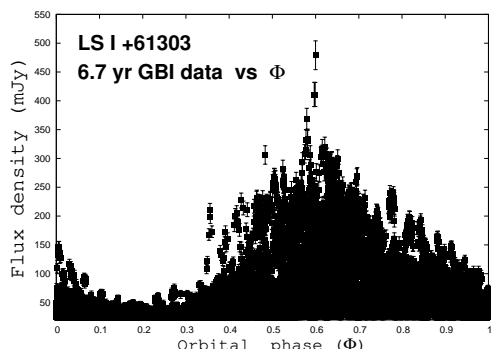}}
\caption{\footnotesize  Radio light curve of LS I +61303 (6.7 yr of Green Bank Interferometer  data) vs orbital phase \citep{massikaufman09}}
\end{figure}
One of the fundamental questions concerning the strong periodic radio
outbursts at  $\Phi\simeq$0.6 of \lsi has been:
why are they  shifted with respect to the periastron passage (i.e.  $\Phi=0.230-0.275$)?

The accretion rate $\dot{M} \propto {\rho_{\rm wind}\over v_{\rm rel}^3}$,
 where $\rho_{\rm wind}$ is the density of the Be star wind and
 $v_{\rm rel}$ is the relative speed between the
accretor and the wind, is
in fact  proportional to the density of the accreted material  \citep{taylor92}.
 The highest
density is obviously  at the periastron.
The explanation for the shift is that the orbit of \lsi is quite eccentric,
e=$0.54-0.7$ \citep{aragona09, casares05}. 
 In an eccentric orbit the different relationship for density and velocity 
(inversely proportional to the power of 3) creates
two  peaks in the accretion
rate curve, one at periastron because of the highest density, 
 and a second one when
the drop in density is compensated by the decrease in velocity towards apastron.
\citet{taylor92} computed the accretion rate curve for different eccentricities and
showed that two peaks begin to appear  for an eccentricity above 0.4.
Whereas the first peak is always toward periastron, the orbital
occurrence of the  second accretion peak
 depends on variations of the wind of the Be star with the period $P_2$=1667~d.
\citet{martiparedes95} computed the accretion rate curve for different wind velocities
and showed that for a stellar wind velocity of
20 km/sec the two peaks become rather close to each other, whereas for a wind velocity
of 5 km/sec they have
an orbital offset of $\Delta \Phi=0.4$. For  periastron
at $\Phi\simeq$ 0.2-0.3 the second peak may therefore appear at $\Phi\sim 0.6-0.7$, as indeed occurs.
\citet{martiparedes95} have shown
that both peaks are above the Eddington limit and therefore one expects that
matter is ejected twice within the 26.5 d interval.
\citet{romero07} applied a smoothed particle hydrodynamics
code to develop  three-dimensional, dynamical simulations for \lsi and found
that indeed the  accretion rate has two peaks per orbit, i.e. a narrow peak at periastron,
and a broad peak that lags the periastron passage by about 0.3 in phase.

Models and simulations predict therefore two peaks. Why do we observe only the second ejection?
\citet{martiparedes95} predicted that near periastron  the ejected relativistic electrons
are embedded in such a strong UV-radiation field
that they  loose their energy by the inverse Compton  (EIC) process: 
no  radio emission but   high energy emission is  predicted.
\citet{boshramon06} computed the inverse compton  losses and the related light curves of emission
for \lsi in the radio band and at high energy. 
\citet{boshramon06} fixed the Be wind parameter to have the second accretion peak at $\Phi$=0.5.
Their results, are here shown in Fig. 2. 
In perfect correspondence to the accretion rate curve 
the radio light curve shows two peaks, one large outburst at $\Phi$=0.5  and
another smaller outburst  at periastron. 
The high energy light curve shows exactly the contrary situation, again
two peaks one at periastron and the other at  $\Phi$=0.5, but the dominant
peak   is at periastron.
During the  second accretion peak the compact object  is at larger distance from  the Be star
 and  therefore inverse Compton losses  are lower:
the associated gamma-ray outburst is weaker and  the electrons  
are able to emit  stronger synchrotron radiation producing the larger
radio outburst.
\begin{figure}
\resizebox{\hsize}{!}{\includegraphics[clip=true, angle=0.]{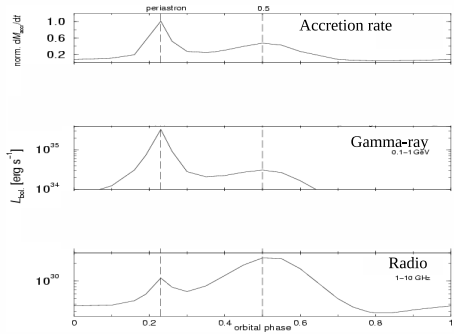}}
\caption{\footnotesize  Computed accretion rate and light curves  by  \citet{boshramon06} (from their Fig. 2). Bosch-Ramon et al. fixed the Be wind parameter to have the second accretion peak at $\Phi$=0.5.
}
\end{figure}
\begin{figure}[]
\begin{center}
\includegraphics[scale=0.27, angle=-90.]{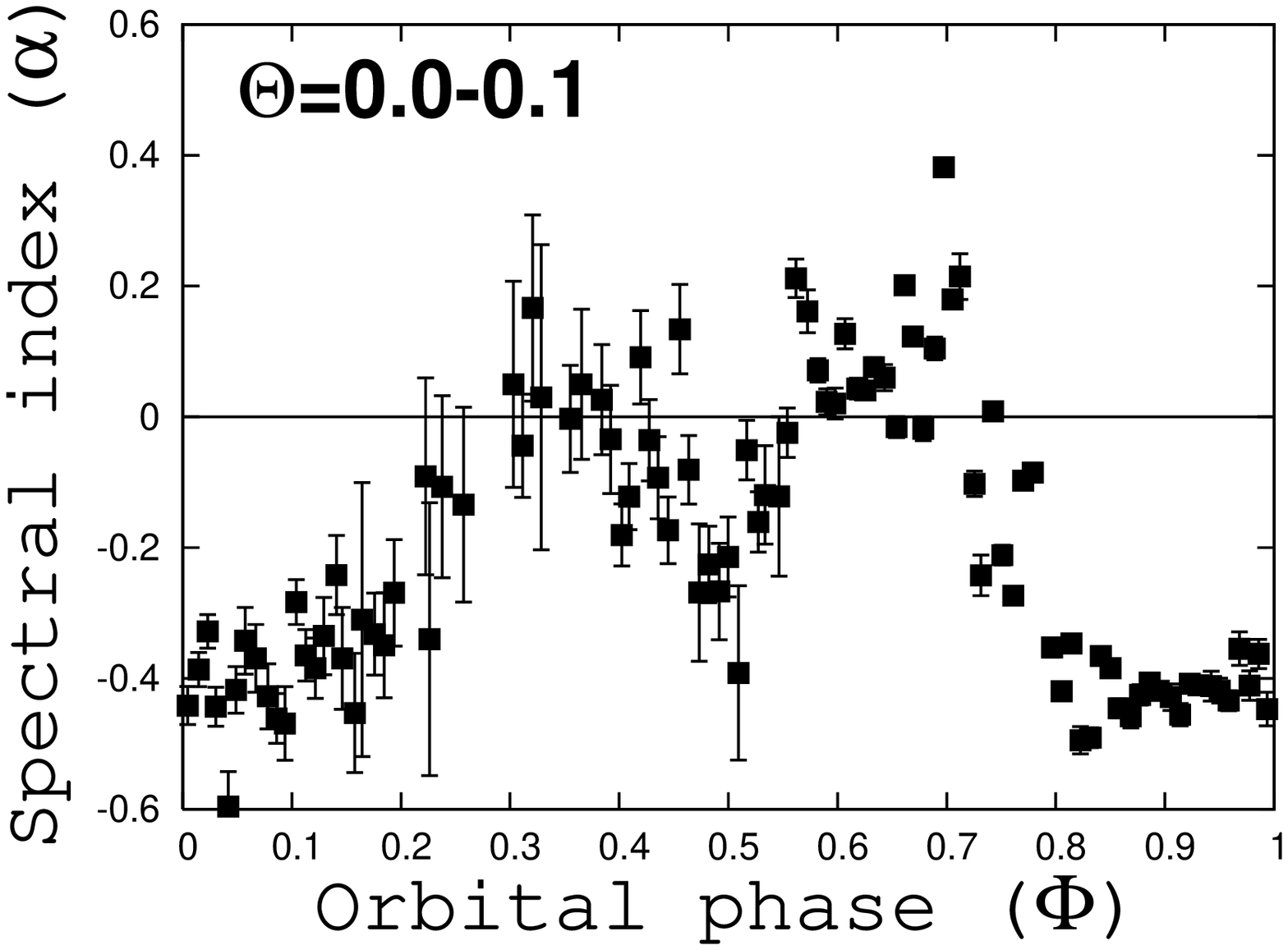}\\
\includegraphics[scale=0.27, angle=-90.]{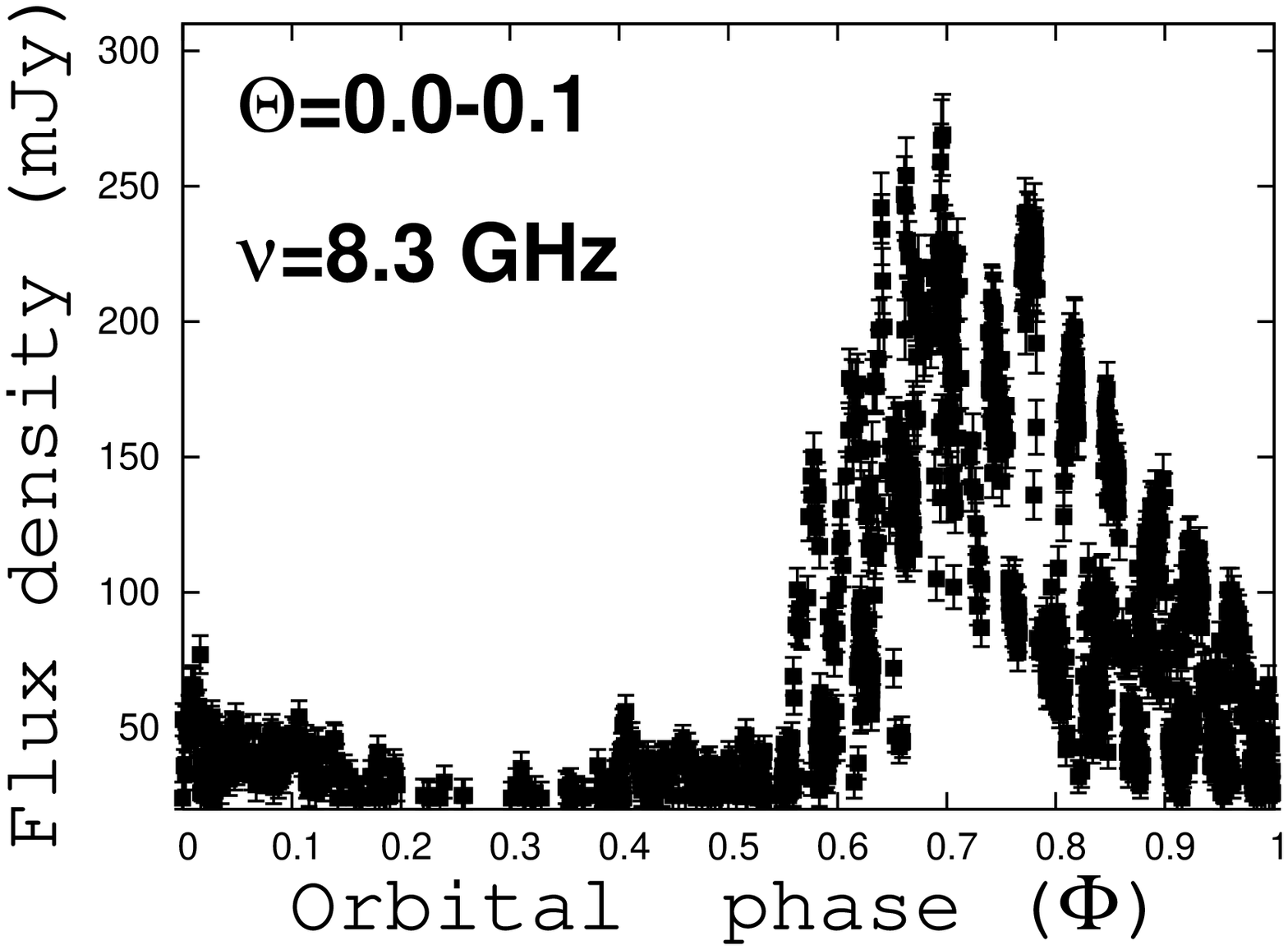}\\
\includegraphics[scale=0.27, angle=-90.]{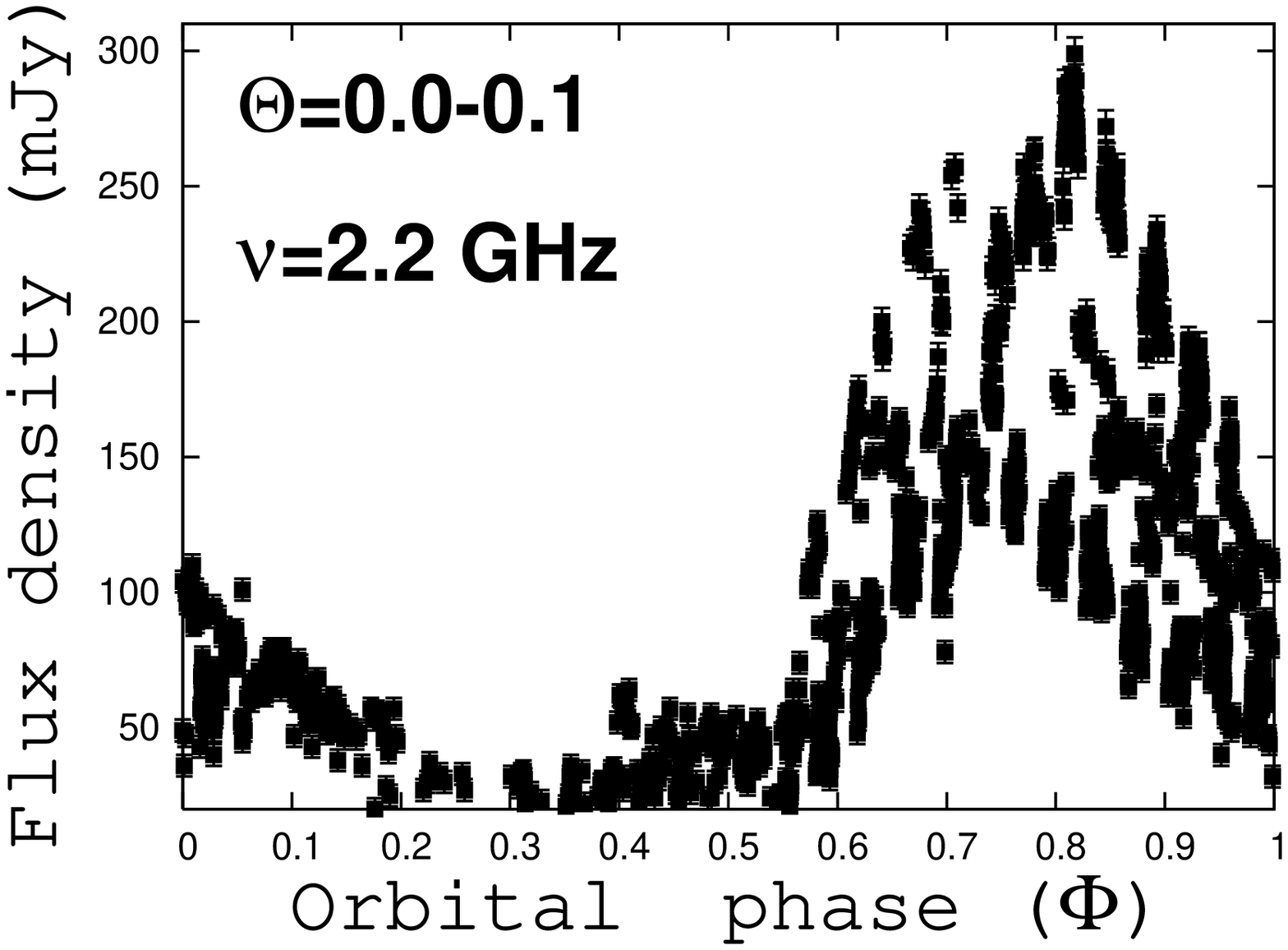}\\
\end{center}
\caption{\footnotesize
\lsp. Spectral index and flux density of Green Bank Interferometer data  
 at 
8.3 GHz and 2.2 GHz  vs orbital phase, $\Phi$,
in the interval $\Theta$=0.0-0.1.
Note that the evolution  from an optically thick to an optically thin spectrum
 occurs twice, giving
 the  $\alpha$ vs $\Phi$ curve  a double-peaked  shape.
\citep{massikaufman09}
}
\label{oo4}
\end{figure}

Gamma-ray observations confirm \citet{martiparedes95} predictions and \citet{boshramon06} calculations.
\lsi was detected by  EGRET   \citep{tavani98}.
As discussed in \citet{massi04},  \citet{massi05} and \citet{massikaufman09}
these data well support  the hypothesis of a high-energy outburst at periastron:
EGRET observations   
at  $\Theta$=0.18 
during a well sampled  full orbit 
show  a clear peak at periastron passage. EGRET observations at   
$\Theta$=0.41, along 
with an increase of the emission again near periastron  
show   even  a second 
   peak at $\Phi\simeq 0.5$ (see Fig. 3 in \citet{massi05}).
 Fermi-LAT observations   \citep{abdo09} were performed  
 at  $\Theta=0.788-0.927$. The Fermi light curve is characterized by a broad peak after periastron 
as well as a smaller peak just before apastron. 
Therefore,  the first gamma-ray peak seems indeed to be persistent,  whereas 
the second gamma-ray peak seems as predicted, to change as function of 
the long term  variations  of the Be star. 

\section{Radio spectral index analysis}
From the previous section we see as the two-peak accretion model, including energetic losses, 
predicts at periastron  a high energy outburst due to
IC  along with a small radio outburst, and associated to the second accretion peak 
a large radio outburst and possibly, depending on the stellar distance, a gamma-ray peak. 
The large radio outburst 
should
follow the typical characteristics of microquasars:
optically thick emission, i.e.  $\alpha\geq 0$ 
(with flux density $S\propto \nu^\alpha$), 
followed by an optically thin outburst, i.e. $\alpha < 0$ 
 \citep{fender04}.
In microquasars the first type of emission, the optically thick radio emission, is related
 to a steady, low velocity, conical  jet centered on the system.
 The following  optically thin outburst, called "transient jet",
is due to shocks caused by the travelling 
of  new highly relativistic plasma,  generated by a transient,
through the underlying,  slow,        
 steady flow 
(Fender et al. 2004; review by Massi in this volume). 
In Fig.3  we see the spectral index and flux density, at 8.3 GHz and 2.2 GHz  vs orbital phase, $\Phi$,
for the GBI data in the interval $\Theta$=0.0-0.1 
(see \citet{massikaufman09} for details). 
Clearly associated with the large outburst of \lsi is
the predicted  evolution for microquasars, from  an optically thick to 
an optically thin spectrum. 
At the bottom of Fig. 3, the light curve at 2.2 GHz reveals that the large outbust at $ \Phi\sim 0.8$ is preceeded by another outburst at  $ \Phi\sim 0.7$.
The spectral index at the top of Fig. 3, 
shows  the different nature of the two outbusts: the minor outburst 
 at $\Phi=0.7$  is optically thick, whereas the larger peak   around
$\Phi=0.8$ is an optically thin outburst.
In particular, the outburst at  $\Phi=0.7$ is related to an  optically-thick-emission
interval
creating a broad peak in the spectral index curve. 
In the context of microquasars this interval corresponds
to the emission from a steady, low velocity conical jet. 
Around periastron in the light curves at 2.2 and 8.3 GHz
one sees only very small, barely detectable outbursts  at  $\Phi\sim 0.3-0.4$.
On the basis of the light curves alone  one would never be able to associate
these negligible peaks to the predicted small radio outburst at
periastron by \citet{martiparedes95} and  \citet{boshramon06}.
When one, however, analyses the spectral index curve at the top of Fig. 3, one sees that
  $\Phi\sim 0.3-0.4$ indeed corresponds again to a 
broad  peak in the spectral index curve.
The evolution  from an optically thick to an optically thin spectrum
 occurs clearly twice giving
the  $\alpha$ vs $\Phi$ curve  a double-peaked  shape, as expected
from  a two-peak accretion curve. 

As shown in Fig. 5 of \citet{massikaufman09}   
this shape is not constant but changes during the long 1667~d cycle  
in agreement with  \citet{martiparedes95} computations for  
variable parameters of the  wind of the Be star.
This variation  of the accretion rate curve/spectral index curve with $\Theta$ 
implies that one can compare
data of different epochs {\it only} when observed in the same $\Theta$ interval.
As an example one sees  that at the top of Fig. 3,
for $\Theta=0.0-0.1$,  the spectral index curve  at $\Phi=0.5$ gives  $\alpha < 0$
(i.e. corresponding to a transient jet); 
at the top of Fig. 4 for $\Theta=0.788-0.927$,  one sees that  
at the same orbital phase $\Phi=0.5$, it results 
$\alpha > 0$ (i.e. corresponding  to a steady slow conical outflow).
We compare here in Fig. 4  the Fermi-LAT gamma-ray curve with  the
spectral index for GBI data
at other epochs than Fermi-LAT observations but in the same  phase
$\Theta=0.788-0.927$.
It is worth noting that both   Fermi-LAT gamma-ray peaks 
correspond to intervals where the radio emission is optically thin.
This correspondence would imply that the electrons responsible for the inverse Compton process creating the gamma-ray emission are those of the
fast transient jet and not those of the 
slow outflow.
\begin{figure}
\resizebox{\hsize}{!}{\includegraphics[clip=true, angle=-90.]{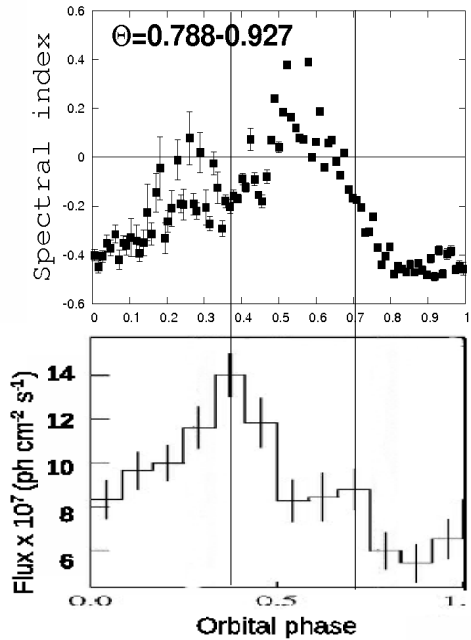}}
\caption{\footnotesize Radio and gamma-ray data observed 
at different epochs but in the same  interval $\Theta$=0.788-0.927 
of the 1667~d periodicity. 
Top: Spectral index of  Green Bank Interferometer data
 at
8.3 GHz and 2.2 GHz.
Bottom: Fermi-LAT data from \citet{abdo09}}
\end{figure}

\section{Conclusions}
We analyzed  the radio spectral index using 6.7 years of GBI radio data of \lsp.
Our main conclusions are that the  periodic ($P_1=26.5$d) large radio outburst
 of \lsi  consists
 of two successive outbursts,
one optically thick and the other optically thin.
This creates one peak in the spectral index curve.
 In  microquasars,
the optically thick emission is associated to a steady jet and the optically thin 
outburst to the transient jet.
We observe that along the 26.5~d orbit
the evolution  from an optically thick to an optically thin spectrum
occurs also around periastron creating a second peak in
the spectral index curve, giving   
the  $\alpha$ vs $\Phi$ curve  a double-peaked  shape.
This result agrees  with the predictions of
the  two peak accretion/ejection model,
with the results of three-dimensional dynamical simulations
 and finally with  gamma-ray data.
All these results indicate  a scenario with a first ejection   around the
 periastron passage 
with low radio emission, but high energy emission,
due to inverse Compton losses caused by the proximity of the B0 star,
and  a second ejection
far away from the Be star, with negligible losses and
therefore with a well observable radio outburst.
Finally, the here given comparison of radio (GBI) and gamma-ray (Fermi-LAT)
data seems to imply that the electrons responsible for the 
inverse Compton process creating the gamma-ray emission are those of the
fast transient jet  and not those of the 
slow  outflow.
\begin{acknowledgements}
I am very  grateful to Marina Kaufman-Bernard\'o and Lisa Zimmermann
for their valuable comments and fruitful discussions.
The Green Bank Interferometer is a facility of the National
Science Foundation operated by the NRAO in support of
NASA High Energy Astrophysics programs.
\end{acknowledgements}

\bibliographystyle{aa}

\begin{thebibliography}{}

\bibitem[Abdo et al.(2009)]{abdo09} Abdo, A.~A., et al.\ 2009,
\apjl, 701, L123

\bibitem[Aragona et al.(2009)]{aragona09}
Aragona, C., et al. 2009, ApJ, 698, 514

\bibitem[Bosch-Ramon et al.(2006)]{boshramon06}
Bosch-Ramon, V., Paredes, J.~M., Romero, G.~E., \& Rib{\'o}, M.\ 2006, \aap, 459, L25

\bibitem[Casares et al.(2005)]{casares05}
Casares, J., Ribas, I., Paredes,
J.~M., Mart{\'{\i}}, J., \& Allende Prieto, C.\ 2005, \mnras, 360, 1105

\bibitem[Fender et al.(2004)]{fender04}
 Fender, R.~P., Belloni, T.~M., \& Gallo, E.\ 2004, \mnras, 355, 1105

\bibitem[Gregory(2002)]{gregory2002}
Gregory, P.~C. 2002, ApJ, 575, 427

\bibitem[Gregory \& Neish(2002)]{gregoryneish02}
 Gregory, P.~C., \& Neish, C.\ 2002, \apj, 580, 1133

%\bibitem[Johnston et al.(2005)]{johnston05}
% Johnston, S., Ball, L., Wang, N., \& Manchester, R.~N.\ 2005, \mnras, 358, 1069 
%\bibitem[Kaufman Bernad{\'o} et al.(2002)]{kaufman02}
%Kaufman Bernad{\'o}, M.~M., Romero, G.~E., \& Mirabel, I.~F.\ 
%2002, \aap, 385, L10

\bibitem[Mart{\'{\i}} \& Paredes(1995)]{martiparedes95}
Mart{\'{\i}}, J., \& Paredes, J.~M. 1995, \aap, 298, 151

%\bibitem[Marscher \& Gear(1985)]{marschergear85} Marscher, A.~P., \& Gear, W.~K.\ 1985, \apj, 298, 114
%
\bibitem[Massi(2004)]{massi04}
Massi, M.\ 2004, \aap, 422, 267

\bibitem[Massi et al.(2005)]{massi05}
Massi, M., Rib{\'o}, M., Paredes, J.~M., Garrington, S.~T., Peracaula, M.,
\& Mart{\'{\i}}, J.\ 2005, High Energy Gamma-Ray Astronomy, 745, 311

\bibitem[Massi \&  Kaufman Bernad\'o(2009)]{massikaufman09}
 Massi, M., \& Kaufman Bernad{\'o}, M.\ 2009, \apj, 702, 1179 

%\bibitem[Mendelson \& Mazeh(1989)]{mendelsonmazeh89} Mendelson, H., \& Mazeh, T.\ 1989, \mnras, 239, 733 

\bibitem[Taylor et al.(1992)]{taylor92}
Taylor, A.~R., Kenny, H.~T., Spencer, R.~E., \& Tzioumis, A. 1992, \apj, 395, 268

\bibitem[Romero et al.(2007)]{romero07}
Romero, G.~E., Okazaki, A.~T., Orellana, M., \& Owocki, S.~P. 2007, \aap, 474, 15

\bibitem[Tavani et al.(1998)]{tavani98}
Tavani, M., Kniffen, D.,
Mattox, J.~R., Paredes, J.~M., \& Foster, R.\ 1998, \apjl, 497, L89

\bibitem[Zamanov \& Mart\'{\i}(2000)]{zamanovmarti00}
Zamanov, R. K, \& Mart\'{\i}, J. 2000, A\&A, 358, L55



\end{thebibliography}

\end{document}